\documentclass[aps,pra,superscriptaddress,twocolumn,10pt]{revtex4-2}
\usepackage{graphicx}
\usepackage{dcolumn}
\usepackage{bm}
\usepackage{nicefrac}
\usepackage{amsfonts,amsmath,amssymb,stmaryrd}
\usepackage{braket}
\usepackage{tabularx}
\usepackage{multirow} 
\usepackage{hhline}
\usepackage{subfigure}  
\usepackage{bbm} 
\usepackage[pdftex]{epsfig}
\usepackage{mathrsfs}
\usepackage{verbatim}
\usepackage{ulem}
\usepackage{array}
\usepackage{cancel}
\usepackage{ifthen}
\usepackage{float} 
\usepackage{listings}
\usepackage{color}
\usepackage{hyperref}
\usepackage{amsmath}
\hypersetup{colorlinks=true,linkcolor=blue,anchorcolor=blue,citecolor=blue,filecolor=blue,urlcolor=blue}

\usepackage[nameinlink,noabbrev]{cleveref}

\begin{document}

\title{Probing the geometry dependence of the Casimir-Polder interaction by matter-wave diffraction at a nano-grating}

\author{Matthieu Bruneau}
\affiliation{Laboratoire de Physique des Lasers, Université Sorbonne Paris Nord, CNRS UMR 7538, F-93430, Villetaneuse, France.}
\affiliation{Leibniz University of Hanover, Institute of Quantum Optics, QUEST-Leibniz Research School, Hanover, Germany.}

\author{Julien Lecoffre}
\affiliation{Laboratoire de Physique des Lasers, Université Sorbonne Paris Nord, CNRS UMR 7538, F-93430, Villetaneuse, France.}

\author{Gabin Routier}
\affiliation{Laboratoire de Physique des Lasers, Université Sorbonne Paris Nord, CNRS UMR 7538, F-93430, Villetaneuse, France.}

\author{Naceur Gaaloul}
\affiliation{Leibniz University of Hanover, Institute of Quantum Optics, QUEST-Leibniz Research School, Hanover, Germany.}

\author{Gabriel Dutier}
\affiliation{Laboratoire de Physique des Lasers, Université Sorbonne Paris Nord, CNRS UMR 7538, F-93430, Villetaneuse, France.}

\author{Quentin Bouton}
\affiliation{Laboratoire de Physique des Lasers, Université Sorbonne Paris Nord, CNRS UMR 7538, F-93430, Villetaneuse, France.}

\author{Thorsten Emig}
\affiliation{Laboratoire de Physique Théorique et Modèles Statistiques, CNRS UMR 8626, Université Paris-Saclay, 91405 Orsay cedex, France.}

%
  
\date{\today}
\begin{abstract}
Atomic diffraction through a nanograting is a powerful tool to probe the Casimir-Polder potential. Achieving precise measurements require simulations to bridge theory and experiment. In this context, we present various approximations and methods of Casimir-Polder potentials, and we analyze their impact on matter-wave diffraction patterns. Our analysis includes the pairwise summation approach, the proximity force approximation, and multiple scattering expansion method. Furthermore, we demonstrate that the influence of Casimir-Polder interactions extends up to 25 nm before and after the nanograting slit, highlighting the importance of accounting for this effect in any accurate analysis.
\end{abstract}

\maketitle
\section{Introduction}

Quantum vacuum fluctuations are characterized by the continuous spontaneous creation and annihilation of virtual photons. In particular, macroscopic surfaces modify these fluctuations, leading to a spatially varying Lamb shift \cite{Buhmann2012}. The gradient of this shift gives rise to an attractive long-range force, known as the Casimir-Polder (C-P) force. This dispersion force, which exhibits a strong dependence on distance at the nanoscale, persists even when atoms are in their ground state. Hence, it plays a crucial role across various fields. For instance, with the emerging of the atom-chip systems as a quantum technology platform, the predominantly attractive nature of the C-P force has been largely perceived as an obstacle for placing atoms close to surfaces \cite{Rahi2010}. This attractive force also impedes the transport of atoms along hollow-core optical fibers \cite{Afanasiev2010} and can lead to the loss of atoms in waveguides utilized for matter waves \cite{Vorrath2010}. Furthermore, C-P forces play an important role for exploring potential modifications to the Newtonian gravitational interaction at the nanoscopic scale \cite{Onofrio2006}. Indeed, the dominance of the C-P force may act to shield these modifications. 

In this context, a range of C-P forces measurements has been performed over the past decade, utilizing various approaches, including the mechanical method, the spectroscopy, the scattering, and diffraction techniques \cite{Laliotis2021}. In this study, we concentrate on the diffraction method where the underlining idea is to let the atoms interact with the slits bars of a transmission nanograting via the C-P potential. The phase shift imprinted by this interaction alters the interference pattern, allowing to extract information related to the C-P potential from the diffraction pattern. This approach has been successfully implemented to various atomic species \cite{Perreault2005, Lepoutre2009, Bruhl2002} and even complex molecules \cite{Brand2015, Simonovic2024}. By combining this process with laser-cooling, slow atomic beam has recently been created, achieving velocities below 20 m/s, and leading to high accuracy and sensitivity to the C-P force. For example, it has recently allowed to distinguish between the non-retarded and the retarded regime in the intermediate range \cite{Garcion2021} or to investigate in details factors such as nanograting geometry, finite-size effects, slit width, opening angles, and patch potentials \cite{Lecoffre2025}. This type of experiment is highly sensitive to various C-P effects, including the exact form of the potential and the effects of the potential both within and outside the slits. Therefore, robust quantitative computations are necessary to connect theory and experiment. 

In this paper, we  investigate the impact of various approximations for the C-P potential  on the diffraction pattern. The studied and compared approximations are the proximity force approximation (PFA) based on the Lifshitz theory, the pair-wise summation approximation, and the recently developed multiple scattering expansion  \cite{Emig2023}. In particular, we analyze the  approximations in the realistic nanograting geometry. Thereafter, by incorporating these approximations into a quantum wave propagation simulation, we study how the C-P force influences the envelope of the diffraction pattern and its effects at the entrance and exit on the atomic wave function.

\section{Matter-wave diffraction experiments description}\label{sec:1}

\begin{figure}
    \centering
    {\includegraphics[width=\linewidth]{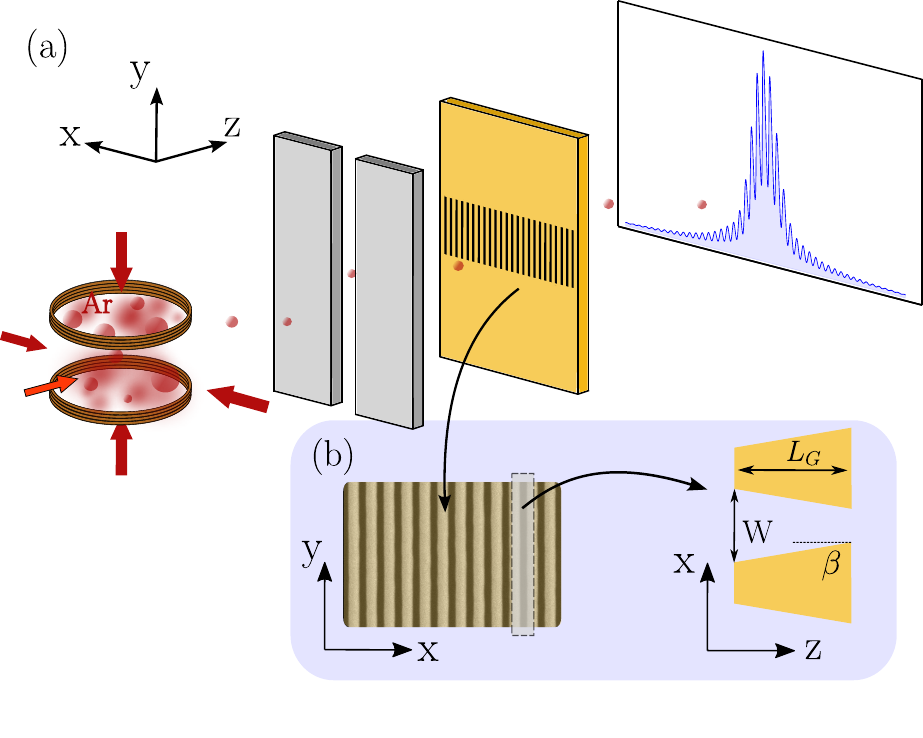}}
    \caption{(a) Argon atoms, trapped in a magneto-optical trap, are initially pushed along the $z$ axis. Prior to their interaction with the nanograting, the angular dispersion of the atomic source is selected by a mechanical slit. The probability density of the atoms are detected and reconstructed after a free propagation of some dozen ms. (b) represents scanning electron microscope image of a standard nanograting utilized in the experiment. The general slit size is trapzoidal, with a slit entrance $W$, a depth $L_{G}$ and a opening angle $\beta$. In this study, we have selected the following parameters: $W=60$ nm, $L_{G}=60$ nm and $\beta=5$°.}
    \label{fig:Exp_set_up}
\end{figure}

Throughout this study and for numerical applications, we will utilize the laser-cooled diffraction experiment described in \cite{Lecoffre2025}, which is illustrated in Fig.~\ref{fig:Exp_set_up}. A detailed description of the experimental setup can be found in \cite{Garcion2021}. In summary, the atomic source on the experiment is composed of many Argon atoms trapped in a magneto-optical trap. The Ar atoms are prepared in the metastable state $^{3}P_{2}$. The wave-functions of these atoms freely expands toward a silicon nitride ($\mathrm{Si_{3}N_{4}}$) transmission grating, featuring nanometric scales. For computational applications, we will use a general form of nanogratring, with commonly used dimensions, as depicted in Fig~\ref{fig:Exp_set_up}. The dimensions selected are a slit size entrance $W$ = 60 nm, a thickness $L_{G}$ = 60 nm  and an opening angle $\beta = 5$°. Employing metastable Argon atoms protects the nanostructure from chemical pollution, while collision effects are avoided due to the low density of the atomic beam. A mechanical slit is strategically placed a few centimeters before the nanograting to control the angular distribution of the source. Measurements are conducted in the far-field regime, where the wave function is reconstructed using microchannel plates in front of a delay line detector. Although the simulations presented in the following sections are based on the aforementioned experiment, the results will remain broadly applicable for many experiments, as most atomic diffraction experiments utilize similar nanograting materials and geometries, and the C-P force is comparable due to similar atomic polarizabilities of the species used.     

\section{Casimir-Polder potential approximations}\label{sec:2}
\subsection{Proximity force approximation (PFA)}
\begin{figure}
    \centering
    \includegraphics[width=\linewidth]{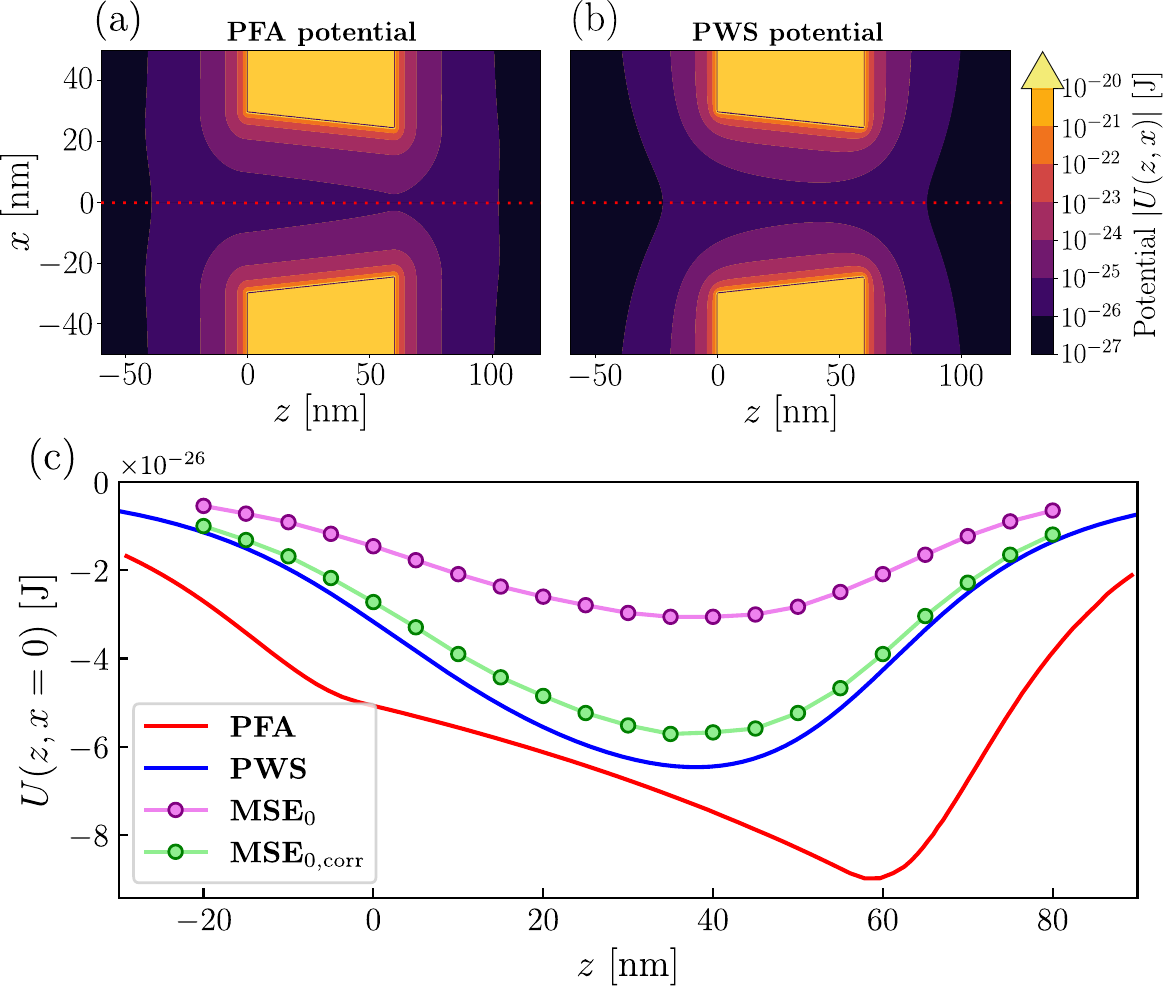}
    \caption{2D Contour plot of the C-P potential with (a) the PFA and (b) the PWS. (c)  C-P potentioal along the symmetry axis of the slit ($x=0$) as given by PFA,  PWS, and 0th order MSE with and without corrections for higher orders.}
    \label{fig:potentials_2D}
\end{figure}

The C-P potential can be computed using the Lifshitz theory, which is valid for an infinite planar surface. For an atom positioned at a normal distance $d$ from the surface, the potential can be expressed as an integral along the imaginary frequency axis \cite{Lifshitz1956}
\begin{align} \label{eqn:C.P_QED}
V_{PFA}(d) &  =  \frac{\hbar \mu_{0}}{8 \pi^{2}} \int_{0}^{+\infty} \mathrm{d}\xi\, \xi^{2} \alpha(i \xi) \int_{0}^{+\infty} \mathrm{d}k_{\parallel}\,\frac{k_{\parallel}}{\kappa^{\perp}}  \nonumber \\
& \times \left( r_{s} - \left( 1+2 \frac{k_{\parallel}^{2}c^{2}}{\xi^{2}} \right) r_{p} \right) e^{-2 d \kappa^{\perp}}
\end{align}
where $\kappa^{\perp} = \left( k_{\parallel}^{2} + \xi^{2}/c^{2} \right)^{1/2}$ is the perpendicular wave-vector, $\alpha(i \xi)$ the atomic polarizability, and $r_{s}$ (respectively $r_{p}$) the $s$-polarized (respectively $p$-polarized) Fresnel-reflection coefficients. Both coefficients write
\begin{eqnarray}
  r_s = \frac{\kappa^\perp -\kappa_{\rm 1}^\perp}{\kappa^\perp +\kappa_{\rm 1}^\perp} \,,\quad r_p = \frac{\varepsilon(i \xi) \kappa^\perp -\kappa^\perp_{\rm 1}}{\varepsilon(i \xi) \kappa^\perp +\kappa^\perp_{\rm 1}}\,,
\end{eqnarray} 
with $\kappa^{\perp}_{\rm 1} = \left( k_{\parallel}^{2} + \varepsilon(i \xi)  \xi^{2}/c^{2} \right)^{1/2}$ and $\varepsilon(i \xi)$ is the relative dielectric function of  $\mathrm{Si_{3}N_{4}}$. Based on the measured optical response \cite{Philipp1973,Luke2015}, the function $\varepsilon(i \omega)$ can be interpolated using
\begin{equation}\label{eq:dielectric_fit}
\varepsilon(i \xi) = 1 + \frac{\varepsilon_{1}}{1+\exp{ \left( \frac{\xi}{\xi_{1}} \right) }} + \frac{\varepsilon_{2}}{1+\exp{ \left( \frac{\xi}{\xi_{2}} \right) }}
\end{equation}
where $\varepsilon_{1} = 7.06$, $\xi_{1}=1.27 \times 10^{14}$ rad/s, $\varepsilon_{2} = 6.00$, and $\xi_{2}=1.64 \times 10^{16}$ rad/s. The Lifshitz formulation takes into account both the retarded and non-retarded regime. In the case of short atom-surface distance (non-retarded limit), it simplifies to $V_{PFA}(d) = C_{3}/d^{3}$ \cite{Buhmann2012}. For an Argon atom in the quantum internal state $^{3}P_{2}$, we obtain $C_{3} = 5.01$ $\mathrm{meV.nm^{3}}$. 
The Lifshitz theory is the basis for the PFA, which has been the core theory for analyzing the majority of precise measurements of C-P forces \cite{Laliotis2021}.
The PFA for the C-P potential for a non-planar surface geometry is  obtained by summing the potential due each position of the two surfaces forming the slit, i.e., of the two columnar bodies adjacent to the slit. Here the distance $d$ is set to the minimal atom-surface separation, measured along the local surface normal. A 2D plot of the potential is presented in Fig.~\ref{fig:potentials_2D}. The PFA is valid only when the atom-surface distance is small, i.e. when the surface is locally well approximated by a semi-infinite planar surface. 

\subsection{Pair-wise summation (PWS)}
However, the PFA based on the Lifshitz theory fails at the edge of the slits, where the surface should not be treated as locally planar. Therefore, in order to more accurately account for the overall geometry of the problem, we also employ the PWS. In this approach, the C-P potential $V_{PW}$ is derived from the summation of the van der Waals potential $V_{vdw} = -C_{6}/r^{6}$ between all the atoms composing the bodies and the Argon atom. Assuming a constant atomic density $\rho$ inside the material, the pair-wise potential reads
\begin{equation} \label{eq:PW_potential}
V_{PWS}=\pm \int_{u}^{\infty} \int_{-\infty}^{+\infty} \int_{0}^{L_G} -\frac{\rho C_6}{r^6} \, dx' \, dy' \, dz'
\end{equation}
where $u=\pm W/2 \mp z' \tan(\beta)$, $r^{2} = (x-x')^2+(y-y')^2+(z-z')^2$, and $\pm$ denotes the upper (+) or lower columnar bodies of the slit (-). This approximation is only valid in the non-retarded limit, meaning for short atom-surface distance, and does not include collective effects inside the materials. An analytic form of Eq.~(\ref{eq:PW_potential}), along with a detailed discussion, can be found in \cite{Lecoffre2025}. To compare with the previous approximation, the pair-wise potential is normalized such as when $L_{G} \rightarrow + \infty$, both potentials coincide in the non-retarded regime. This condition imposes $C_{3}= \pi \rho C_{6} / 6$. A 2D plot of the pairwise potential $V_{PWS}$ is depicted in Fig.~\ref{fig:potentials_2D}. As anticipated, both potentials are in excellent agreement near the slit center ($z=L_{G}/2$) and close the slits walls, with a relative difference of less than $1 \%$. Nevertheless, at the center ($z=L_{G}/2$ and $x=0$), the two approximations differ by $8 \%$. At this position,  retardation effects contribute  $13\%$ of the total potential, while the effect of finite size of the slits begins to manifest. At both the slit center and edge ($z=0$ and $x=0$), the accuracy of the two approximations reduces due to the growing contribution of the retarded regime and to the intricate local geometry,  which cannot be approximated as locally planar, leading to a relative difference of $50 \%$. 

\subsection{Multiple scattering expansion (MSE)}
The two previous models are uncontrolled approximations for the C-P potential, and hence their validity throughout the entire spatial domain is unknown. An exact computation of the C-P potential for non-planar, non-spherical, and non-cylindrical geometries poses significant challenges. Here, to approach this problem systematically, we use the recently developed general multiple scattering expansion method introduced in \cite{Emig2023} to infer the C-P at the slit center $x=0$. This approach is based on fluctuating electric and magnetic surface currents. We shall compare the MSE result to the two previous approximations. 

First, we briefly review the MSE for the C-P potential of an atom in the vicinity of dielectric bodies with permittivities $(\epsilon_1$, $\mu_1)$, surrounded by a medium with  permittivities $(\epsilon_0$, $\mu_0)$. A detailed description and derivation can be found in~\cite{Emig2023,Bimonte2023}. Here, we can assume that the atom has an isotropic (scalar) electric polarizability $\alpha$. At zero temperature, the C-P potential for an atom at position ${\bf r}_0$ can be expressed as an integral along the imaginary frequency axis ($\kappa=\xi/c$), 
\begin{align}\label{eq:MSE}
& V_{MSE}({\bf r}_0) = -2\,  \hbar c  \int_0^\infty d \kappa \, \alpha(i\xi) \, \kappa \, \int_S ds_{\bf u} \int_S ds_{{\bf u}'}\,\\ \nonumber
&  \sum_{p,q\in \{E,H\}}\!\!\!\!\! \text{tr} \left[  \mathbb{G}_0^{(Ep)}({\bf r}_0,{\bf u}) \left[(\mathbb{I}-\mathbb{K})^{-1}\right]^{(pq)}({\bf u},{\bf u}') \mathbb{M}^{(qE)}({\bf u}',{\bf r}_0)\right]
\end{align}
where the integrals extend over all positions of the  surfaces $S$ of the dielectric bodies (here the nano-grating), tr denotes the trace of the $3\times 3$ matrix, and ${\mathbb G}_0$ the free Green tensor for a homogeneous space filled by a medium with permittivities $(\epsilon_0$, $\mu_0)$.
Here $\mathbb{K}$ is a surface scattering operator, defined for two surface points ${\bf u}$, ${\bf u}'$ by
\begin{equation}
\label{eq:operator_K_general}
\mathbb{K}({\bf u},{\bf u}') = 2 \mathbb{P} (\mathbb{C}^{i}+\mathbb{C}^{e})^{-1} {\bf n}({\bf u}) \times  \left[ \mathbb{C}^{i}  \mathbb{G}_1({\bf u},{\bf u}')- \mathbb{C}^{e} \mathbb{G}_0({\bf u},{\bf u}')\right] 
\end{equation}
and $\mathbb{M}({\bf u},{\bf r})$ is the bulk-surface operator
\begin{equation}
\label{eq:operator_M}
\mathbb{M}({\bf u},{\bf r}) =  -2 \mathbb{P} (\mathbb{C}^{i}+\mathbb{C}^{e})^{-1} \mathbb{C}^{e}\, {\bf n}({\bf u}) \times  \mathbb{G}_0({\bf u},{\bf r}) 
\end{equation}
which describes the wave propagation from a point ${\bf r}$ in the  space outside the bodies to the first point ${\bf u}$ on the surface $S$. ${\bf n}({\bf u})$ is the outward unit normal vector to the surface $S$, $\mathbb{G}_1$ is the free Green tensor for a homogeneous medium with permittivities $(\epsilon_1$, $\mu_1)$ (see App.~E of \cite{Bimonte2023} for explicit expressions for $\mathbb{G}_0$ and $\mathbb{G}_1$), and the matrices
\begin{equation}
\mathbb{P} = \big(\begin{smallmatrix} 0 & -1 \\ 1 & 0 \end{smallmatrix} \big), \quad \mathbb{C}^{i}={\rm diag}(\epsilon_1,\mu_1)\, , \quad
\mathbb{C}^{e}={\rm diag}(\epsilon_0,\mu_0) 
\end{equation}
act in the 2-dimensional polarization space $(E,H)$. For the nano-grating we set $\epsilon_0=\mu_0=\mu_1=1$ and $\epsilon_1$ equals the dielectric function of $\mathrm{Si_{3}N_{4}}$.

The inverse of $\mathbb{I}-\mathbb{K}$ in Eq.~(\ref{eq:MSE}) cannot in general be computed in closed form. The multiple scattering expansion avoids direct inversion by expanding the inverse in powers of $\mathbb{K}$ where each additional power introduces an additional surface integral over $S$. Here, we consider the lowest order, corresponding to replacing $\left[(\mathbb{I}-\mathbb{K})^{-1}\right]^{(pq)}({\bf u},{\bf u}')$ by $\delta({\bf u}-{\bf u}')\delta_{pq}\mathbb{I}_{3\times 3}$, so that only a single surface integral over $S$ remains to be computed, and also the first order for some distance to check convergence.  We can further assume that the nano-grating is infinitely extended in the $y$-direction. Due the resulting translational invariance along the $y$-axis the computation of the surface integrals is more easily performed by expressing the Green tensors in terms of the partially Fourier transformed scalar Green function $g_\sigma({\bf r},{\bf r}')=\int_{-\infty}^\infty \!\!\frac{d q_y}{(2\pi)^2}e^{i q_y(y-y')} K_0(\sqrt{(x\!-\!x')^2+(y\!-\!y')^2}\sqrt{q_z^2\!+\!\epsilon_\sigma\mu_\sigma \kappa^2})$ where $K_0$ is a Bessel function (see App.~E of \cite{Bimonte2023} for the relation between the Green tensors and the scalar Green function). Due to this representation the C-P potential is given by integrations over $\kappa$ and $q_y$ and 1-dimensional surface integrals in the $xz$-plane. The resulting C-P potential to lowest order MSE is shown in Fig.~\ref{fig:potentials_2D}. Here we restrict the MSE to this order to get a first estimate of the C-P potential, and leave the explicit inclusion of higher orders and a detailed convergence check for a longer work. However, we employ a simple scheme to extrapolate from lowest order MSE result all higher orders: The exact C-P potential for the (within the slit region) closely related geometry  of an atom between two infinite, parallel flat surfaces (2PS) is known exactly \cite{Buhmann2012}. Hence, we can check easily the convergence rate for this geometry by comparing the first two MSE orders to the exact result. We find  for the 2PS geometry that (1) the 1st order MSE contribution is reduced by $0.82$ relative to the 0th order MSE contribution  to the C-P potential, and (2) the sum of the 0th and 1st order is only reduced by $0.98$ compared to the exact result, demonstrating very fast convergence. A full computation of the 1st order MSE at $z=30$nm for the actual slit geometry shows that the 1st order MSE contribution is reduced by $0.84$ relative to the 0th order MSE contribution, and is hence consistent with the convergence for the simpler 2PS geometry. For this 2PS geometry the 0th order MSE contribution is $0.53$ of the exact result. Hence we use this factor to correct for higher MSE orders in the actual slit geometry. This corrected C-P potential is also shown in Fig.~\ref{fig:potentials_2D}. It shows a similar dependence on $z$ than the PWS, with a difference of $12.5\%$ at $z=30$nm which can be understood from non-additivity effects and the absence of retardation effects in the PWS estimate. \\\\

\section{Impact of C-P potential approximations on the resulting diffraction pattern}\label{sec:3}
\subsection{Matter-wave diffraction model}
The PFA and PWS, which can be evaluated using minimal computational resources across the entire spatial domain under consideration, are incorporated into a theoretical model of matter-wave diffraction. The wave-propagation model is based on the time-dependent Schrödinger equation and goes beyond the standard semi-classical approach. This model was introduced in our previous work \cite{Garcion2024} in one dimension (along the $x$-axis). In this study, we extend the model to two dimensions (along the $x$ and $z$ axis). Along the y-axis, the nano-grating slits are sufficiently long so that the C-P potential can be regarded as constant in this direction, as well as the diffraction pattern along this direction. In the following paragraph, we briefly describe the method, emphasizing its novel aspects. Further details regarding this theoretical approach can be found in \cite{Garcion2024}. As illustrated in Fig.~\ref{fig:Exp_set_up}, the diffraction pattern is characterized by the interfringe spacing, peak size, and envelope. The interfringe spacing and peak size result from the presence of multiple slits and the properties of the atomic source, while the envelope, which is the signature of the C-P potential, arises from a single slit. Hence, we only focus on the envelope of the diffraction signal, reducing the simulations to the propagation through a single slit. The initial wave function at the entrance of the slit is a plane wave propagating along the $z$-axis at a velocity $v$, modulated by a Gaussian envelope with a standard deviation of $\sigma=1$nm along the same axis. The time-dependent Schrödinger equation is solved numerically using the second-order split-operator technique. A mask function is included in the simulations to model losses when atoms hit the surface and to prevent numerical reflections. The mask function is a continuous sigmoid with a characteristic slope of $0.1$nm. Finally, the 2D stationary phase approximation is used to propagate the wave function from the exit of the slit to the detector. With the detector positioned at a fixed location $z_{0}$, the final signal is obtained by integrating the projected wave function over the $z$-axis, see Fig.~\ref{fig:inside_slits}. We verified that such a procedure is equivalent to performing a time integration. In all simulations performed, the initial velocity is assumed to be $v=15$ m/s. The spatial resolution of the propagation grid inside the slit is approximately $\delta x \simeq $ 10 pm, $\delta z \simeq$ 70 pm and $\delta t = 1.0$ps. These values ensure numerical convergence of the simulations (in both amplitude and phase). The computational grid contains $2^{14} \times 2^{11}$ points, with a higher resolution along the $x$ axis. A different grid was used to study the propagation up to $30$nm before and after the slit. In this case, the resolution was $2^{14} \times 2^{12}$ points and the parameters were $\delta x \simeq $ 10 pm, $\delta z \simeq$ 40 pm and $\delta t = 0.5$ps. The distance between the nano-grating and the detector is $D=250$mm, ensuring that the simulations are in the far-field regime. 

\subsection{Influence of the C-P potential inside the slits}

\begin{figure}
    \centering
    \includegraphics[width=\linewidth]{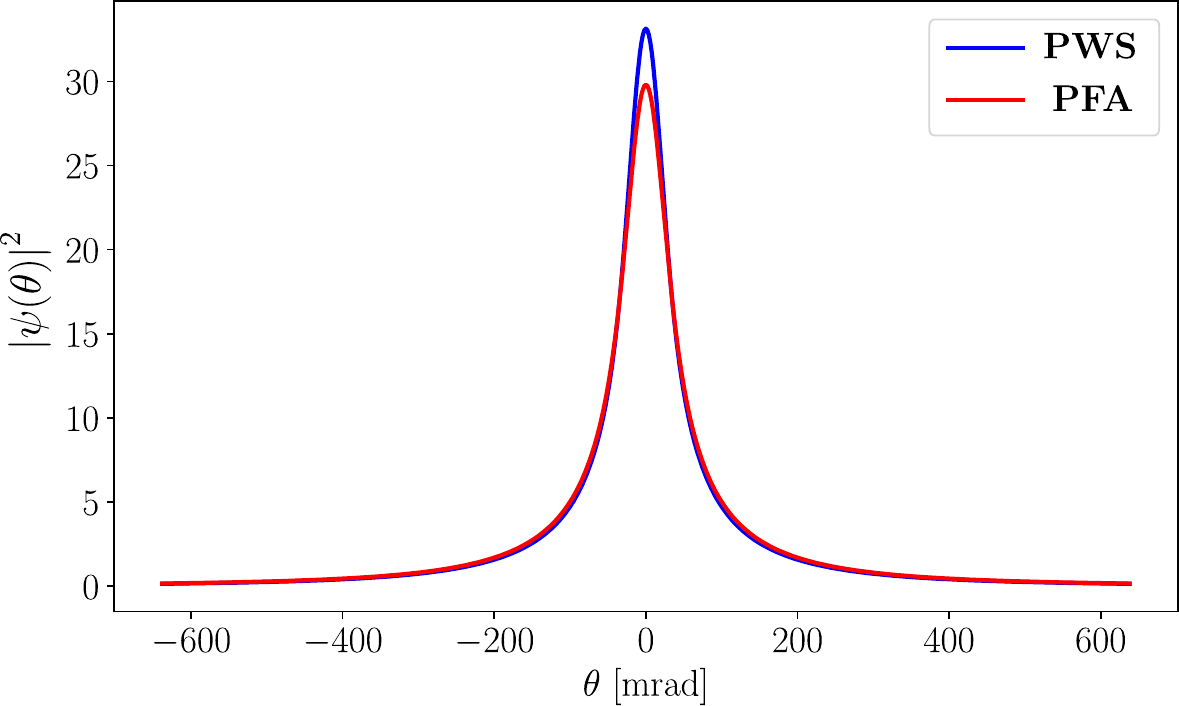}
    \caption{Integrated probability density along the $z$ axis of the single-slit wave function, derived from the 2D simulation, as a function of the diffraction angle $\theta = \mathrm{tan}(x/D)$. $D$ is the distance from the grating to the detector. The blue curve corresponds to the probability density obtained from the PWS, whereas the red curve illustrates the probability density calculated using the PFA.}
    \label{fig:inside_slits}
\end{figure}

First, we concentrate on the impact of the C-P potential on the diffraction pattern, taking into account only the regions within the slits. In Fig.~\ref{fig:inside_slits},  the results of the 2D propagation simulations for the PFA and PWS are shown. We observe that the envelope signal differs according to the chosen approximation. The PFA overestimates the potential inside the slits, resulting in an envelope that is larger than the one produced by the PWS, as illustrated in Fig.~\ref{fig:inside_slits}. In order to quantify this difference, we assess the $C_{3}$ coefficient in the PWS which provides the closest diffraction pattern generated by the PFA. We thus introduce the quantity  
\begin{equation} \label{eq:indicator}
A = \frac{\displaystyle\int \left|\big|\psi_{PFA}( \theta)\big|^2-\big|\psi_{PWS}(\theta)\big|^2\right| d \theta }{\displaystyle\int \big|\psi_{PFA}(\theta)\big|^2\, d \theta}\,.
\end{equation}
where $\theta$ is the diffraction angle, while $\big|\psi_{PFA}( \theta)\big|^2$ and $\big|\psi_{PWS}( \theta)\big|^2$ are the probability densities on the detector inferred using the PFA and the PWS, respectively. This quantity measures the relative difference between the two models. A smaller value of $A$ indicates that the two diffraction patterns are very similar. By minimizing $A$, we determine that $C_{3} = 6.3 \,\mathrm{meV nm^{3}}$, which is $1.27$ times larger than the strength coefficient of the PFA. This difference emphasizes the importance of precisely computing the C-P potential. We also conduct the same study using the 1D wave propagation simulation, applying the equation $x=vt$, i.e., neglecting the force along the propagation axis $z$. This transformation allows us to reduce the 2D problem to a 1D problem, both time dependent. We observe similar results as in the 2D simulation. We attribute this similarity to the small value of the  angle $\beta$, which renders the system nearly symmetric and the Hamiltonian close to being separable. 

\subsection{Influence of the C-P potential at the entrance and exit of the slits}
A thorough comparison between theory and  results of competitive C-P force measurements from matter-wave diffraction necessitates a careful attention to all phase shifts induced by the atomic wave function. In particular, the C-P potential causes phase shifts in the atomic wave functions not only inside the slits, but also before and after the slits itself. If not included in the theoretical calculation, this additional phase shift can lead to significant errors in the C-P potential measurements derived from the diffraction pattern. Such quantification of error is thus crucial for using the C-P effect as a new test for extensions to the Standard Model for example \cite{Mostepanenko2008}. 

\begin{figure}
    \centering
    \includegraphics[width=\linewidth]{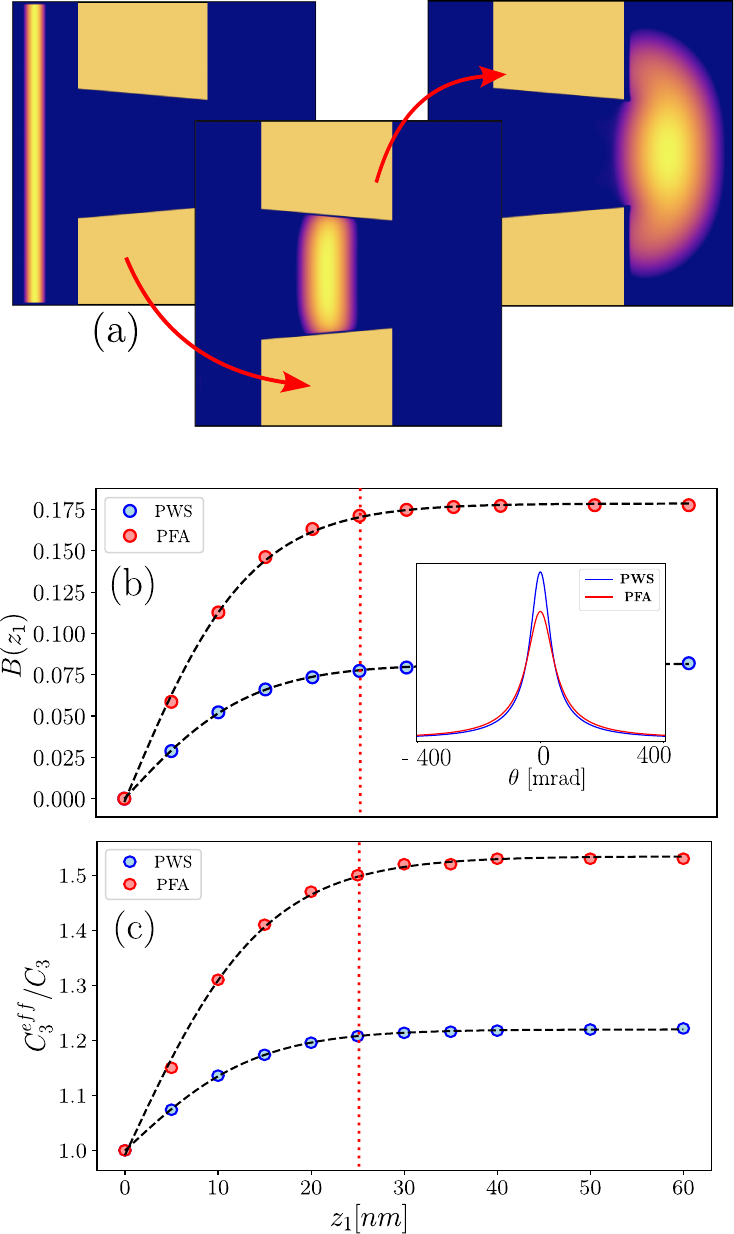}
    \caption{(a) 2D typical time evolution of the wave packets shown in a logarithmic colormap. The  wave packet is initiated at a distance $z_1$ before entering the slit and propagates until it reaches the same distance $z_1$ after exiting the slit. (b) Evolution of the parameter $A$ as a function of $z_1$. Here, $A$ is a comparison between the density probability $|\psi(z_{1})|^{2}$ obtained from simulations starting at $z=z_1$ and the density probability $|\psi(z_{1}=0)|^{2}$ from simulations beginning at $z_1=0$. Blue dots represent the results from propagation simulations conducted using the PWS, while the red dots correspond to simulations performed with the PFA. The black dashed lines represent a fit that serves as a guide for the eye. Inset: density probability at the detector for both potentials at $z_1$ = 25 nm. (c) Evolution of $C_3^{eff}/C_3$ as a function of  $z_1$. The vertical red dotted line indicates the point when both $A$ and $C_3^{eff}/C_3$ reach 95\% of their maximum value, corresponding to $z_1 \simeq$ 25 nm.}
    \label{fig:before_after_slits}
\end{figure}

We initiate the wave propagation simulations at a distance $z_{1}$ prior to the entrance of the slits. By symmetry, the simulations end at the same distance $z_1$ after the exit of the slits (refer to Fig.~\ref{fig:before_after_slits}). By varying the distance $z_1$, we investigate its effect on the envelope of the diffraction pattern. The simulations are conducted using the same parameters as in the previous section, i.e. with the same nano-grating geometry and an incoming velocity set at $v=15$ m/s. We perform the simulations using the PFA and PWS in both 1D and 2D. We do not observe differences between the 1D and 2D simulations, due to the symmetry of the problem. Furthermore, we implement absorption boundary conditions at the edges of the simulation domain to eliminate any numerical reflections. To assess the influence of $z_{1}$ on the diffraction pattern, we introduce the parameter $B(z_1)$, which is analogous to the parameter $A$ and is defined as follows :
\begin{equation} \label{eq:indicator_B}
B(z_1) = \frac{\displaystyle\int \left|\big|\psi( \theta,z_1)\big|^2-\big|\psi(\theta,z_1=0)\big|^2\right| d \theta }{\displaystyle\int \big|\psi(\theta,z_1=0)\big|^2\, d \theta}\,
\end{equation}
where $\big|\psi( \theta,z_1)\big|^2$ is the probability density on the detector using PFA or PWS approximations. The results are illustrated in Fig.~\ref{fig:before_after_slits}. We observe that the parameter $B(z_1)$ saturates at $z_{1}=25$ nm for both approximations, indicating that within the range $0$ nm $< z_{1}<25$ nm, the contribution of the C-P potential remains significant and must be accounted for. Interestingly, both potentials yield the same boundary. Nevertheless, the saturation value of $B(z_1)$ differs from both models: it reaches $B=0.082$ for the PWS, while it reaches $B=0.178$ for the PFA. This indicates that the PFA has a greater influence on the final wave function compared to the PWS. To provide a more quantitative analysis, we also study the error in the C-P potential approximations as a function of the distance $z_1$ for both models. For each distance $z_1$, we identify the closest strength coefficient $C_{eff}$ on the diffraction pattern relative to the scenario where $z_1$=0 nm, characterized by the $C_3$ coefficient. Our findings reveal that the errors arising from neglecting the entrance and the exit of the slits can reach up to 22$\%$ for the PWS and 53$\%$ for the PFA, as shown in Fig.~\ref{fig:before_after_slits}. This finding emphasizes the necessity of incorporating the potential beyond the slits and highlights the need for precise computations of the C-P potential in the future, given that the results are highly dependent on the approximation used. This is particularly crucial because the information derived from the measured diffraction pattern relies on the underlying approximation.

\section{Conclusion}
In this paper, we have presented the effect of different approximations for the C-P potential on the diffraction pattern. By only considering regions within the slits, we find a quite significant difference between the PFA and PWS. Besides, we study the influence of the approximations for the C-P potential at the entrance and exit of the slits. Our findings indicate that simulations have to be conducted at least up to $z_1=25$ nm both prior to and following the slits. This result remains consistent across both approximations. However, the final results strongly depends on the approximations, emphasizing the need for precise computations of the C-P potential, which we do at the center of the slit using a multiple scattering expansion. Extending this calculation to the entire nanograting geometry will open the door to precise comparison between the experiment and the theory. This analysis is a perquisite for any future metrology investigations, such tests of gravity at short distances \cite{Antoniadis2011}.

\acknowledgments
We acknowledge the financial support from the French National Research Agency (Grant No. ANR-24-CE30-4859-01). 


\end{document}